# Unraveling the stable cathode electrolyte interface in all solid-state thin-film battery operating at 5V

Ryosuke Shimizu[1+], Diyi Cheng[2+], Minghao Zhang[1,*], Bingyu Lu[1], Thomas A. Wynn[1], Randall Burger[1], Min-cheol Kim[1], Guomin Zhu[1], and Ying Shirley Meng[1,2,3,*]

[1]*Department of NanoEngineering, University of California San Diego, La Jolla, CA, USA.*
[2]*Materials Science and Engineering Program, University of California San Diego, La Jolla, CA, USA.*
[3]*Pritzker school of Molecular Engineering, University of Chicago, Chicago, IL, 60637, USA*
[+] Equally contributed.
* Co-correspondence: miz016@ucsd.edu, shirleymeng@uchicago.edu

**Abstract:** Spinel-type $LiNi_{0.5}Mn_{1.5}O_4$ (LNMO) is one of the most promising 5 V-class cathode materials for Li-ion batteries that can achieve high energy density and low production costs. However, in liquid electrolyte cells, the high voltage causes continuous cell degradation through the oxidative decomposition of carbonate-based liquid electrolytes. In contrast, some solid-state electrolytes have a wide electrochemical stability range and can withstand the required oxidative potential. In this work, a thin-film battery consisting of a LNMO cathode with a solid lithium phosphorus oxynitride (LiPON) electrolyte is tested and their interface before and after cycling is characterized. With Li metal as the anode, this system can deliver stable performance for 600 cycles with an average Coulombic efficiency > 99%. Neutron depth profiling indicates a slight overlithiated layer at the interface prior to cycling; a result that is consistent with the excess charge capacity measured during the first cycle. Cryogenic electron microscopy further reveals intimate contact between LNMO and LiPON without noticeable structure and chemical composition evolution after extended cycling, demonstrating the superior stability of LiPON against a high voltage cathode. Consequently, we propose design guidelines for interface engineering that could accelerate the commercialization of a high voltage cell with solid or liquid electrolytes.

**Introduction**

Li-ion batteries (LIBs) are dominant battery technologies for portable electronic devices and electrical vehicles due to their high energy density, thermal stability, and long cycle life[1–4]. However, a wider adoption of LIBs requires gravimetric energy densities in excess of 350 Wh kg$^{-1}$ (1 Wh = 3,600 Joules) at the cell level, and up to 500 Wh kg$^{-1}$ for more than 1000 cycles[5,6]. At present, the energy densities of mass-produced LIBs are limited to 200 Wh kg$^{-1}$ - 250 Wh kg$^{-1}$ at the cell level[7–10]. The cathode is regarded as a critical component in improving the capacity of commercial cells. With a high operating voltage (4.7 V vs. Li$^+$/Li$^0$)[11–14], a spinel-type cathode material LiNi$_{0.5}$Mn$_{1.5}$O$_4$ (LNMO) could potentially empower industrial producer to achieve these high energy density goals. Recently, the strong desire to eliminate cobalt in cathode materials has sparked a renewed interest in this class of oxides[15]. Various attempts to fabricate LNMO/graphite batteries that exhibit high voltage, relatively high energy density, and fast charging capabilities using organic liquid electrolyte have been carried out worldwide[16–26], but they all suffered from excessive degradation and limited cycle life, especially when stored or cycled at a highly charged state[27]. The primary reason is that most common liquid electrolytes (e.g., carbonic ester solvent combinations with lithium hexafluorophosphate (LiPF$_6$) solute) are prone to oxidization and subsequent decomposition on the cathode surface as a cell's voltage rises over 4.5 V during charging. This is caused by a lack of an effective passivation layer[27–31], which helps prevent decomposition of the liquid electrolyte, and can cause to a cell to continuously degrade.

All-solid-state batteries (ASSBs) may provide a viable pathway to use LNMO and achieve the desired high energy density and cycling stability. ASSBs have received enormous attention over the last few decades as they have good intrinsic safety, high packing density, and a relatively large electrochemical stability window with the potential to enable both high voltage cathodes and metallic lithium anodes[32,33]. A key factor in successful integration of LNMO in ASSBs will be pairing the electrode material with a compatible electrolyte. Solid state electrolytes (SSEs) prevent catalytic dissolution of transition metals from the cathode into the electrolyte[28,34], which leads to capacity loss and graphite anode degradation when liquid electrolytes are used[28,34–37]. Lithium phosphorus oxynitride (LiPON) is one of the most promising candidates for this application as it has a wide electrochemical stability window up to 5.5 V[38], a modest ionic conductivity ($\approx 10^{-6}$ S cm$^{-1}$)[38], suitable mechanical properties[39,40], and has demonstrated cycling stability against LNMO cathodes and lithium metal anodes[41,42]. The prevailing form of LiPON material as a thin film synthesized by physical vapor deposition provides an ideal platform for investigating interfaces against highly oxidative/reductive electrodes.

LiPON as a solid-state electrolyte has been studied with various types of cathode materials, including LiCoO$_2$ (LCO)[43–45], LiMn$_2$O$_4$[46,47], and LiNi$_{0.6}$Mn$_{0.2}$Co$_{0.2}$O$_2$ (NMC622)[48]. Santhanagopalan et al. observed a lithium accumulation between LCO and LiPON upon repeated cycling, which impacts the ion transport at the interface and counts for irreversible capacity loss[43]. Later Wang et al. identified the cause of lithium accumulation as a disordered LCO formation at the LCO/LiPON interface, which continues to grow at an elevated temperature and leads to a further performance degradation[44,45]. Phillip et al. reported that NMC622/LiPON interface is stable at high voltage (4.5 V vs. Li$^+$/Li$^0$), but the cell suffers from degradation of the bulk cathode due to amorphization, resulting in severe capacity loss[48]. All these research efforts shed light on possible degradation modes at the cathode/electrolyte interface. However, most of these effects are caused by structural change(s) within the cathode and not from LiPON decomposition. This contrasts starkly with the degradation mechanisms occurring in liquid analogues. In the case of LNMO system, Li et al. demonstrated a LNMO/LiPON/Li thin-film full cell that cycled over 10,000 cycles with 90.6% capacity retention[41]. This exemplary cycling performance between 3.5 V and 5.1 V suggests a stable interface between LiPON and the high voltage cathode. Nevertheless, the underlying mechanism that provides such exceptional stability remains elusive, largely due to a lack of available characterization tools that can access the buried interfaces and tackle the air-/beam-sensitivity of LiPON[49].

Meanwhile, tremendous research efforts have also been invested in the study of the cathode electrolyte interface (CEI) in liquid electrolyte systems, which may hint at the potential origin of the LNMO/LiPON interface stability. The interfacial phenomena in liquid electrolyte systems have been widely characterized through spectroscopic and microscopic methods. X-ray photoelectron spectroscopy (XPS)[20–24,27] and attenuated total reflectance Fourier transform infrared spectroscopy (ATR-FTIR)[22,27] were employed to identify the chemistry of the CEIs. After cycling with a carbonate-based electrolyte, a CEI layer forms on the LNMO surface which includes decomposition products of $LiPF_6$ solute and the organic solvents LiF, $Li_2CO_3$, $Li_xPO_yF_z$ and polymerized ethylene carbonate (PEC). Continuous electrolyte decomposition occurrs due to the non-uniformity of the CEIs and an insufficient passivation effect. Cryogenic electron microscopy (cryo-EM) has shown that an uneven CEI layer can form after 50 cycles within a conventional carbonate electrolyte. Improved cycling and a uniform CEI was observed by the same technique but for a sulfone-based electrolyte[26]. Another factor to be considered is the addition of binders and conductive agents in composite cathodes for liquid electrolyte systems. These additives result in parasitic reactions with an electrolyte and can make deconvolution of interfacial reactions between active materials and electrolytes challenging. As such, an ideal interface between LNMO and LiPON must limit electrolyte decomposition, result in conformal CEI formation, minimize cathode structural change, and be free of conductive agents.

An all-solid-state thin film format was employed in this study to examine the origin of the stable interface between LNMO and LiPON. The samples consisted of dense electrode layers without binder, conductive carbon, or coating materials. Neutron depth profiling (NDP) was utilized to delineate the lithium concentration profile across the LNMO/LiPON interface. Results from this measurement were coupled with first-principles computation and cryogenic electron microscopy (cryo-EM) to investigate interfacial chemistries and textures. Based on these findings, crucial characteristics of the solid electrolyte that impact the interfacial stability are discussed and a proposal is made of key factors that facilitate the design of stable, high voltage cells by rational interface engineering.

## Results and discussion
### *Electrochemical behavior of LNMO/LiPON/Li full cell over long-term cycling*

The presence of a stable interface was first demonstrated by a thin film battery consisting of a high-voltage spinel LNMO cathode, a LiPON solid electrolyte, and a lithium metal anode. The detailed architecture of the cell is shown in **Figure S1**, where the deposited LNMO thin film displays a well crystallized structure with (111)-plane-orientated texture in **Figure S2 (a)**. This is in good agreement with the films produced by Xia et al.[50,51] **Figure 1** illustrates the cycling performance of the full cell. As shown in **Figure 1 (a)**, the 1st cycle charge profile displays an excess capacity compared with the subsequent cycles, while the voltage profiles at the 2nd cycle and 600th cycle resemble each other, implying an irreversible reaction during the 1st cycle and a superior cyclability afterwards. The cycling stability is further demonstrated by a Coulombic efficiency of ≈ 99.6% for 600 cycles in **Figure 1 (b)**. Compared with the LNMO cathode performance in liquid electrolyte such as **Figure S2 (b)**, two characteristic features are observed in the thin film battery with the LiPON solid electrolyte: *i*) an excess capacity during the first charge and *ii*) cycling stability observed for the course of 600 cycles. Note 1 mAh is equal to 3.6 C (SI units).

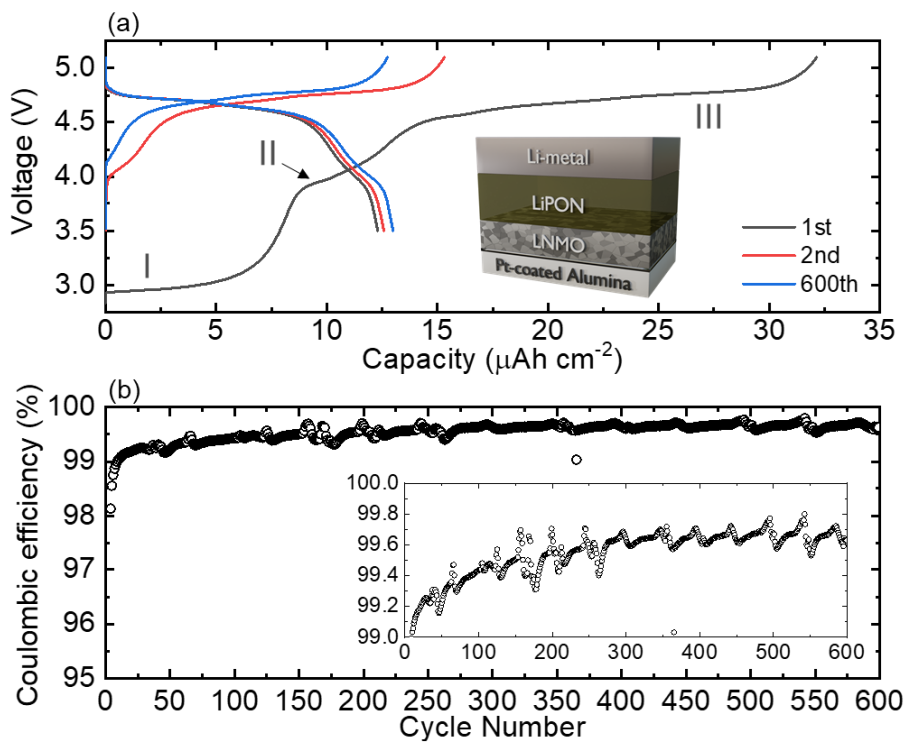

**Figure 1. Electrochemical performance of LNMO/LiPON/Li full cell.** (a) Voltage profiles of LNMO/LiPON/Li thin film full cell at 3.5 V - 5.1 V at the 1st, 2nd, and 600th cycle. (b) Coulombic efficiency of the cycled cell, which was charged/discharged at a about C/7 for the 1st and 2nd cycles, and 4C for the rest of cycles. The C rate was set back to C/7 at the 600th cycle. Uncertainty of the areal capacity in the voltage profile is ~3%, attributed to the estimation of the cell active area.

The voltage profiles indicate that the dominant redox reactions in the LNMO cathode are 4.7 V vs. $Li^+/Li^0$ (shown as plateau region III in **Figure 1 (a)**) and correspond to the $Ni^{2+}/Ni^{3+}$ and $Ni^{3+}/Ni^{4+}$ redox couples during the initial charging process[52,53]. A small amount of $Mn^{3+}/Mn^{4+}$ redox couple occurring near 4 V (plateau region II in **Figure 1 (a)**) contributes to the nominal capacity when part of the Mn species in a pristine LNMO thin film has an oxidation state lower than 4+. The LNMO thin film is likely overlithiated chemically during the full cell fabrication, which leads to a lower Mn oxidation state in the surface region of LNMO. This is apparent in regions I and II (**Figure 1 (a)**), where an excess of charge capacity during the first cycle is caused by a plateau at 2.9 V and 4 V. Note that the pristine LNMO thin film cycled in liquid electrolyte does not exhibit an overlithiated feature as indicated by **Figure S2 (b)** and previous studies[50,51]. The overlithiation of LNMO is proposed to be caused by LiPON deposition and chemical reactions between LNMO and LiPON during or after the deposition.

*Li concentration gradient across LNMO/LiPON interface*
Neutron depth profiling (NDP), is a robust method by which select light elements (e.g., Li) can be quantified[54]. Unlike XPS depth profiling technique, NDP is a nondestructive approach,[55,56] which enables the quantification of the average Li concentration through the sample along the thickness direction. There have been numerous studies over the last several years, where NDP has been applied to the study of Li-ion battery materials[54,57–60]. In this study, NDP was utilized to examine the Li concentration profile of a pristine LNMO/LiPON sample to investigate the possibility of LNMO overlithiation.

**Figure 2 (a)** is a schematic of NDP measurement setup, where samples are mounted in a vacuum chamber and cold neutrons are directed at a sample. Along the pathway, neutron may react with $^6$Li and generates charged particles, $^4$He$^{2+}$ (alpha) and $^3$H$^+$ (triton), with characteristic kinetic energies. Note that alpha particles are filtered out by the thin polyamide film cover and only the triton particles are detected. The energy loss of the charged particles as they pass through the material is due to the stopping power of the material. Li concentration as function of depth is obtained by plotting the detected number of triton particles as a function of final kinetic energy. During this experiment, a LNMO/LiPON thin film sample was measured by NDP, with a LNMO thin film measured as a reference for the interface sample. **Figure 2 (b)** exhibits triton-based Li areal concentration depth profiles from the LNMO/LiPON sample (black), LNMO sample (red), and their subtraction (blue) in Li atoms cm$^{-2}$. The subtracted profile represents variations in Li across the bulk and interface regions of sample after the LiPON layer is added to the LNMO.

To estimate the contributions from the bulk LiPON and interface regions, the subtracted curve (blue curves in **Figure 2 (b) and (c)**) is fit with a Weibull function (details explained in **supporting information** and **Table S1**)[61] in the energy range between 2551 keV and 2358 keV, where bulk LiPON is dominant in the LNMO/LiPON sample. The fit model is then extrapolated to the lower energy region to estimate the Li concentration contributed from pure LiPON (magenta curve in **Figure 2 (c)**). Lastly, the subtraction of the extrapolated model (magenta curve in **Figure 2 (c)**) from the calculated curve (blue curve in **Figure 2 (c)**) is plotted as a green line in **Figure 2 (c)**, which represents the interfacial effect from the LNMO/LiPON sample. As shown in the inset of **Figure 2 (c)**, a noticeable difference in Li concentration between the subtracted (blue) and the extrapolated curve (magenta) can be observed. The positive difference (green) suggests a slight increase of Li concentration at the interface between LNMO and LiPON. Such Li concentration increase (atoms cm$^{-3}$) in LNMO is further estimated by integrating areal Li concentration in atoms cm$^{-2}$ from the energy range between 2358 keV to 2200 keV, which falls into LNMO region. The integration gives a Li concentration increase of ≈ 4.00×10$^{20}$ atoms cm$^{-3}$ at the LiPON/LNMO interface, which is ca. 3% of the concentration compared with the designed stoichiometry of LiNi$_{0.5}$Mn$_{1.5}$O$_4$ cathode.

NDP results indicate an increased Li concentration at the LNMO/LiPON interface. One possible source of this increase is overlithiation of the LNMO surface. Different lithiation states of LNMO were therefore investigated with first-principles calculations to further understand the impact of overlithiation on LNMO material. The $L_xNMO$ modeling structure was altered from $x = 0$ to $x = 2$ to represent the delithiated state and overlithiated state of LNMO, respectively. (**Figure 2 (d)**). **Figure 2 (e)** displays the average magnetizations of Mn and Ni species calculated at each lithiation state; this can be used as an indicator of the oxidation state(s) of transition metals[62,63]. The magnetization number of Ni decreases from 1.7 $\mu_B$ to 0.5 $\mu_B$ (1 $\mu_B$ = 9.274×10$^{-24}$ J T$^{-1}$, SI units) during the delithiation (charging) process ($x = 1$ to 0). This indicates that the Ni redox changes from $Ni^{2+}$ to $Ni^{4+}$ for the charging process[62,63]. The magnetization number of Ni does not vary due to overlithiation ($x = 1$ to 2). In contrast, Mn magnetization changes from 3.2 $\mu_B$ to 3.7 $\mu_B$ when the structure is overlithiated, suggesting a Mn reduction as more Li is inserted into LNMO.

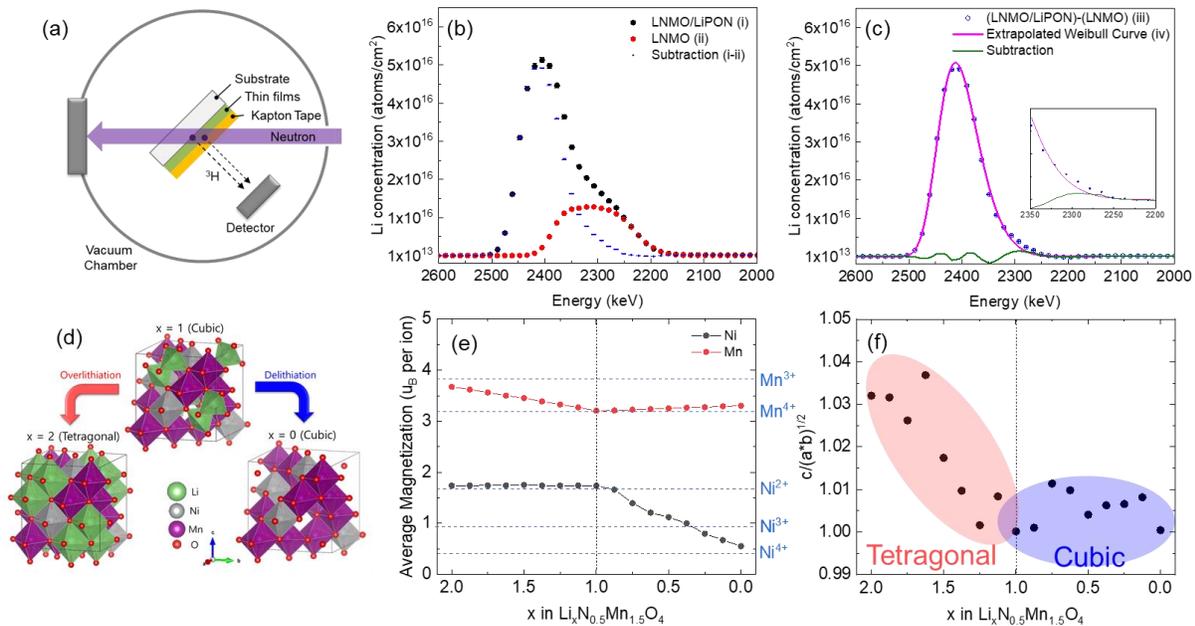

**Figure 2**. **Li concentration across LNMO/LiPON interface.** (a) Schematic image of neutron depth profiling (NDP) setup. (b) Li concentration profile (atoms/cm$^2$) of LNMO/LiPON (black), bare LNMO (red), and subtraction of bare LNMO from LNMO/LiPON as a function of detected triton energy. (c) Li concentration profile of subtraction of bare LNMO from LNMO/LiPON (blue, same data in (b)), Li concentration of bare LiPON simulated with Weibull function (magenta), and subtraction of simulated curve from (LNMO/LiPON)-(LNMO) curve (green). (d) Structures of $Li_xNi_{0.5}Mn_{1.5}O_4$, which are used for DFT calculation, at $x = 1$ (pristine), $x = 0$ (delithiated), and $x = 2$ (overlithiated). (e) Averaged magnetization of Ni (black) and Mn (red) at each lithiated state of LNMO. (f) Relative c length of the supercell at each lithiation state from $x = 0$ to $x = 2$ in $L_xNMO$.

Along with the change in magnetization numbers, the relative lattice parameter evolution of the supercell at each lithiation state is plotted in **Figure 2 (f).** When $x$ in the $L_xNMO$ structure exceeds 1, the lattice parameter in $c$ direction becomes relatively longer than the $a$ and $b$ directions. This leads to a distortion of the lattice in the $c$ direction and, consequently, a phase transformation to a tetragonal phase. Such Jahn-Teller (J-T) distortion is attributed to the presence of $Mn^{3+}$. The Mn cation is normally located at octahedral site bonded with O anions. This causes the classical ligand field splitting (crystal field splitting) and, as a result, d-electron states can be split into triplet ($t_{2g}$) and doublet ($e_g$) states, which stabilizes the Mn-O bonding configuration. In the case of $Mn^{3+}$, three of the four 3d electrons are in $d_{xy}$, $d_{yz}$ and $d_{zx}$ orbitals in the $t_{2g}$ state and leaves one $d_{z2}$ electron in $e_g$ state (high spin state). The Mn-O octahedral structure becomes elongated in the $c$ direction to minimize Coulombic repulsion between the electrons in Mn $d_{z2}$ orbital and the electrons in O 2p orbital as the $d_{z2}$ electron orbitals are parallel to the *c* direction of the octahedron. Therefore, the presence of J-T distortion when $x > 1$ implies the reduction of Mn from 4+ to 3+. This matches with the average magnetization results in **Figure 2 (e)**.

To further verify the oxidation states of the transition metals, the partial density of states (PDOS) of Mn at $x = 1$ (lithiated), and $x = 2$ (overlithiated) are plotted in **Figure S3 (a)** and **(b)**, respectively. PDOS of $L_{1.0}$NMO in **Figure S3 (a)** shows that all the d-orbital electrons below the Fermi level occupy the up-spin state and correspond to an electronic configuration of $Mn^{3+}$ ions. In contrast, $L_{2.0}$NMO in **Figure S3 (b)** has a new occupied state under the Fermi level at around -1 eV. This indicates $Mn^{4+}$ is reduced to $Mn^{3+}$ at $x = 2$ through the addition of one electron in d-orbital. These computational results predict the redox behavior of Mn species when LNMO is overlithiated, well correlating the excess charge capacity from the prolonged plateau at 2.9 V (**Figure 1 (a)**).

*Mn oxidation state evolution at LNMO/LiPON interface*
Scanning transmission electron microscopy/electron energy loss spectroscopy (STEM/EELS) were then conducted to experimentally verify the oxidation state of Mn across the LNMO/LiPON interface. Cryogenic protection was applied on the TEM specimens during STEM measurements, given the beam sensitivity of LiPON as reported in the past study[49]. **Figure 3 (a)** displays the cryo-STEM high-angle annular dark field (HAADF) image of the pristine LNMO/LiPON interface, where LNMO and LiPON regions can be identified based on differences in contrast. The highlighted spots represent the region where EELS spectra were extracted. Mn L-edge spectra are plotted in **Figure 3 (b)**. The low signal intensity detected in the Mn L-edge spectrum at spot 0 indicates the absence of Mn in this region and that the major component is LiPON. The characteristic peaks of Mn L-edge appear in the spectrum from spot 1 to spot 5, suggesting the range of the LNMO/LiPON interface region. The $L_3$ and $L_2$ peaks of Mn L-edge correspond to electron transition from $2p_{3/2}$ and from $2p_{1/2}$ to unoccupied d orbitals, respectively[64]. The oxidation states of the transition metal species can be identified by analyzing the intensity ratio between $L_3$ and $L_2$ peaks and comparing it to the values collected on standard Mn compounds in which the oxidation states of Mn are known. A step function is commonly used to fit these spectra (**Figure S4**) so that the intensity ratio of $L_3$ peak to $L_2$ peak can be calculated and plotted[64], as shown in **Figure 3 (c)**. In **Figure 3 (c)**, Mn at spot 1 shows a $L_3/L_2$ intensity ratio ≈ 2.9, while the ratio decreases to ≈ 2.4 for the Mn from spot 2 to spot 5. According to Wang et al., $L_3/L_2$ intensity ratio of 3.85, 2.6, and 2.0 corresponds to $Mn^{2+}$, $Mn^{3+}$, and $Mn^{4+}$ oxidation states[64], respectively. Therefore, the oxidation state of Mn at the starting of the LNMO/LiPON interface is likely to be between $Mn^{3+}$ and $Mn^{2+}$. This Mn oxidation state at the interface was more reduced than the one found within the LNMO bulk, which is comparable with a bare LNMO thin-film sample (**Figure S5**). The computational results reveal that Mn oxidation state is reduced from 4+ to 3+ when LNMO is overlithiated and form $L_x$NMO ($x > 1$), as shown in **Figure 2 (d)**. The combination of Mn L-edge intensity ratios, computational results and electrochemistry indicate that Mn at LNMO/LiPON interface (spot 1) shows an oxidation state of 3+, which is likely attributed to the overlithiation of LNMO, while bulk LNMO shown at spot 2 to spot 5 has mixed states of $Mn^{3+}$ and $Mn^{4+}$. Another possible cause of Mn reduction is the bombardment of LiPON during sputtering. Song et al.[52] has shown that oxygen vacancy can trigger $Mn^{3+}$ generation from $Mn^{4+}$ for charge compensation, and cathodes with more oxygen vacancy can deliver a longer plateau at 4.0 V in the voltage profile. The $Mn^{3+}$ ions should be oxidized to $Mn^{4+}$ during the charge in the 1$^{st}$ cycle and remain above 3+ after the discharge to 3.5 V as the electrochemical lithiation voltage corresponding to $Mn^{4+} \rightarrow Mn^{3+}$ is at 2.8 V[53,65].

**Figure 3 (d)-(f)** shows the Mn oxidation states at LNMO/LiPON interface after 500 cycles. Compared to the pristine interface (**Figure 3 (c)**), the cycled interface has much decreased $L_3/L_2$ intensity ratio at spot 1 in **Figure 3 (f)**, indicating a less reduced Mn environment. During the charging process in the 1st cycle, excess lithium atoms that caused overlithiation of the LNMO are extracted, a process manifests as a plateau at 2.9 V in the electrochemical measurement. Reduced Mn ($Mn^{3+}$) due to overlithiation and LiPON deposition is also oxidized in the 1st charge, rationalizing the measured increased Mn oxidation state in subsequent cycles.

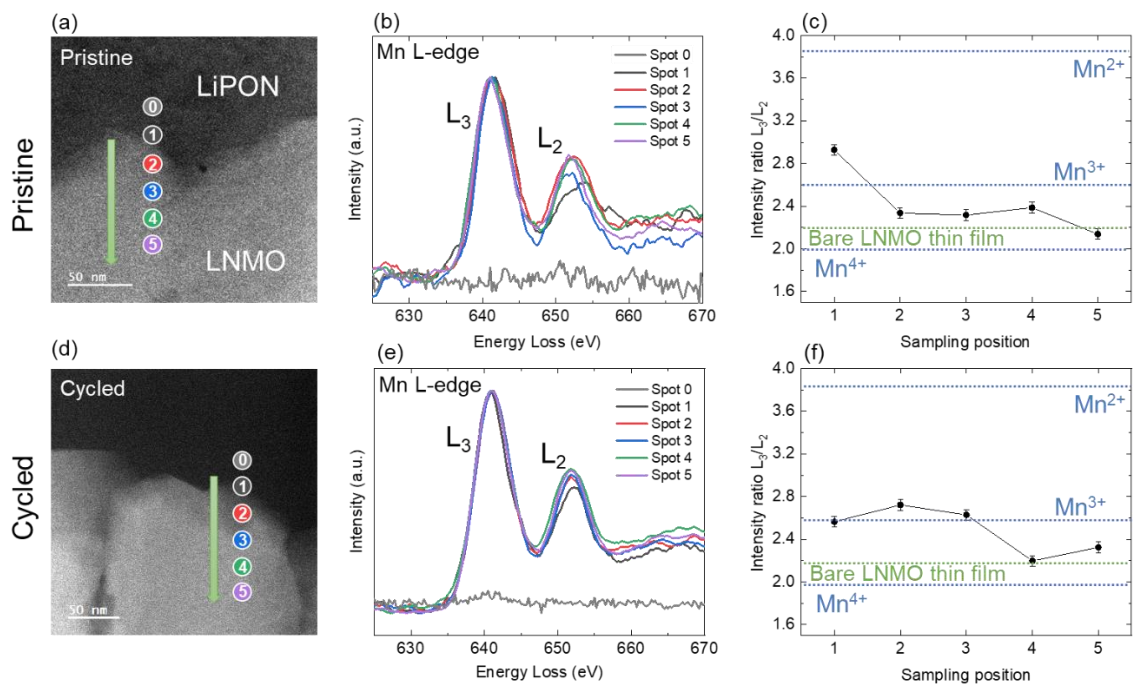

**Figure 3. Mn oxidation state changes across the pristine and cycled interface.** (a-c) Analysis of pristine LNMO/LiPON interface and (d-f) cycled LNMO/LiPON interface. Cryo-STEM image (a) at pristine interface and (d) at cycled interface. (b) Mn L-edge EELS spectra at each point of pristine interface and (e) cycled interface. Intensity ratio of Mn $L_3$ peak to $L_2$ peak with error bars (c) for pristine interface and (f) cycled interface. Details of intensity integration method are illustrated in Figure S4.

*Nanostructure and morphology at LNMO/LiPON interface*
The nanostructure and morphology of the layers were additionally characterized to elucidate the interfacial stability and to compliment the above discussed chemical analyses. Cryo-TEM images at the LNMO/LiPON interface before and after cycling were obtained and summarized in **Figure 4**. **Figure 4 (a)** shows the high-resolution TEM (HRTEM) image of pristine LNMO/LiPON interface. No voids or cracks are formed when LiPON is deposited on LNMO. Detailed nanostructure was also inspected and shown in **Figure 4 (b) and (c)**. Lattice fringes observed in these images match well with (111) lattice plane of LNMO[66,67] and indicates that LNMO maintains its crystal structure after the full cell preparation. A cycled LNMO/LiPON interface is shown in **Figure 4 (d)**. It displays surprisingly intimate contact between the LNMO and LiPON layers and an absence of voids or cracks, even after 500 cycles. **Figure 4 (e)** and **(f)** demonstrate that the crystal structure of LNMO is maintained in both the bulk and interface regions. Although the cell was cycled at a rate of 5 C, which can be considered as fast cycling and results in the cell having 40% of its original capacity, the LNMO/LiPON interface remains intact and shows no signs of structural change or defect formation. In **Figure 4 (g)-(k)**, a clear difference between crystalline region and amorphous region is highlighted by the yellow curve along the cycled LNMO/LiPON interface. Very little CEI is observed in this region. These observations provide good evidence for the long-term structural stability of the LNMO/LiPON interface.

An interesting phenomenon was observed in the LiPON after 500 cycles. Cryo-HRTEM imaging of the LiPON in a pristine LNMO/LiPON sample (**Figure S6 (a)**) indicates nanocrystal formation in areas highlighted by the yellow dot lines. These nanocrystalline species are identified as $Li_3PO_4$ and $Li_2PO_2N$ through the features in Fast Fourier transform (FFT) patterns shown in **Figure S6 (b)** and **(c)**. These are likely to be decomposition products of LiPON based on past computational results[68]. Such partial decomposition occurs only at a small portion of the interface. Similar phenomenon is observed at the cycled LNMO/LiPON interface (**Figure S7** with lattice patterns displayed in (**a**) and FFT analysis displayed in (**c)-(e)**)). The presence of LiPON partial decomposition in both pristine and cycled LNMO/LiPON interfaces implies that such decomposition is driven chemically instead of electrochemically. This is discussed further in the following section.

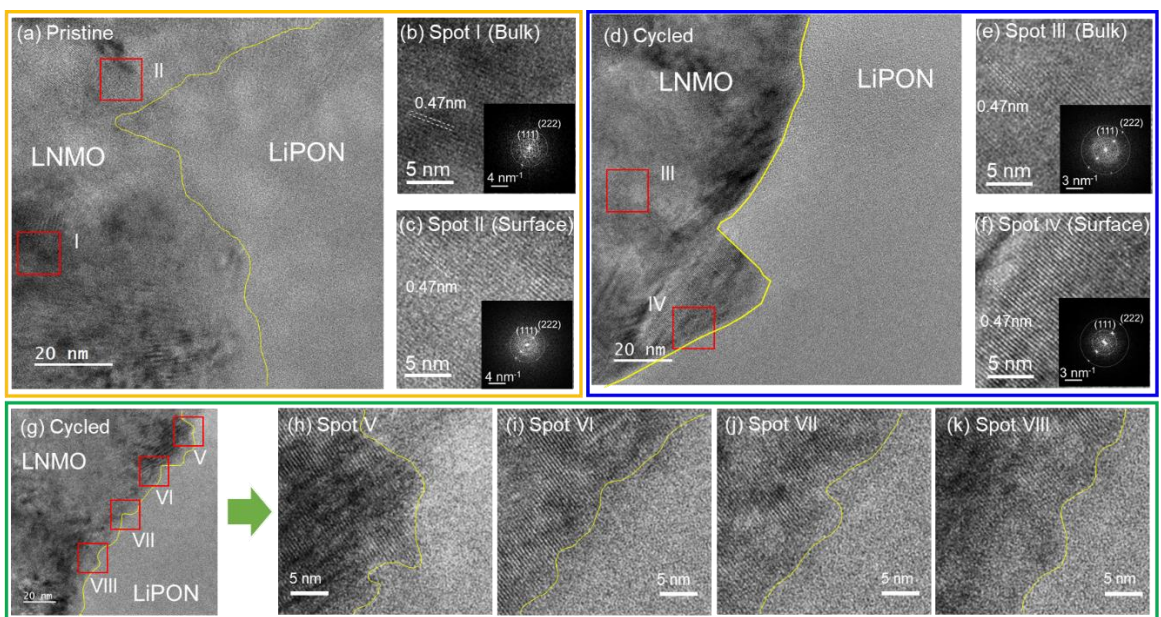

**Figure 4. Interfacial nanostructure and morphology at the LNMO/LiPON interface** (a) Cryo-TEM image of pristine LNMO/LiPON interface and (b) zoomed LNMO images of bulk (inset; FFT image) and (c) surface (inset; FFT image). (d) Cryo-TEM image of cycled LNMO/LiPON interface and (e) zoomed LNMO images of bulk (inset; FFT image) and (f) surface (inset; FFT image). (g) Another cryo-TEM image of cycled LNMO/LiPON interface and (h-k) zoomed interface images at different spots.

### *An electrochemically stable and mechanically compatible solid-solid interface*
The stability at LNMO/LiPON interface stems from two separate perspectives: *i*) the intrinsic structural resilience of LNMO, and *ii*) the unique characteristics of LiPON. The LNMO material's spinel crystal structure determines its ability to accommodate excess Li. Upon overlithiation, LNMO experiences a phase transformation from cubic to tetragonal structure while it homes extra

Li in the structure without sabotaging the cyclability and interface stability. Spinel analogue anode material, $Li_4Ti_5O_{12}$ (LTO), undergoes a similar phase transformation when extra Li is inserted into the structure while maintaining LTO's highly reversible cycling performance.[69] Contrary to spinel oxides, layered oxides are prone to unfavorable changes with excess Li. LCO suffers from the formation of a disordered phase that comprise of a mixture of $Li_2O$ and $CoO$ between ordered LCO and LiPON during sputtering process. This results in an extra interfacial impedance and can inhibit cycling performance.[43–45] $Li_xNi_{0.8}Mn_{0.1}Co_{0.1}O_2$ (NMC811), another layered cathode material, experiences a 45.5% capacity loss in subsequent cycles after deep discharging to 0.8 V vs. $Li/Li^+$ in the initial cycle in the liquid electrolyte system.[70] The ability of spinel oxides to withstand surface overlithiation renders LNMO suitable for coupling with LiPON electrolyte, and serves to increase the surface Li chemical potential of the cathode for better interfacing with solid electrolytes.

Three aspects can be considered regarding the uniqueness of LiPON material. The overarching factor is the electrochemical stability of LiPON. Partial decomposition of LiPON is observed after its deposition onto LNMO surface. This indicates a certain extent of chemical incompatibility. However, long-term cycling does not result in additional decomposition of LiPON. Rather, most of the LiPON remains amorphous along the interface. This implies that LiPON is electrochemical stable against a high voltage LNMO cathode. Further EELS analysis at the cycled LNMO/LiPON interface shows no noticeable changes in the P and N chemical environments, which suggests the stability of LiPON's chemistry against highly oxidative potential (**Figure S8**). Such electrochemical stability could be influenced by the increased Li concentration estimated from the NDP results. The ascending Li content across the interface from LNMO to LiPON region represents a gradually ramping chemical potential, which mitigates the oxidative potential from high voltage LNMO and protects LiPON from further decomposition[32]. Coupling with the analogous chemical gradient feature observed at Li/LiPON interface elsewhere[49], the presence of gradient interfaces, which exhibits exemplary electrochemical stability, hint on the plausibility of engineering solid-solid interfaces so that drastic chemical potential differences between electrode and electrolyte can be diminished to achieve desired stability.

Another aspect of LiPON being stable against LNMO ties into its mechanical characteristics. Herbert et al. measured the Young's modulus of LiPON by nanoindentation and obtained a value of 77 GPa[39]. This high modulus provides LiPON a rigid characteristic. However, Lee et al. reports a 11% - 13% volume change of $L_xNMO$ cathode upon lithiation from x = 0 to x =2.[53] Given the polycrystalline feature of LNMO cathode, anisotropic stress generated at the interface during such repeated volumetric changes can cause catastrophic impact on LiPON's bulk structural integrity such as cracking or delamination. , while none of these issues were observed to occur at the LNMO/LiPON interface. The fact that LiPON retains an intact contact with LNMO after long-term cycling suggests that the LiPON in this study may be a relatively soft material. This is consistent with some observations on the flexibility of LiPON in literature[40,71], yet further experimental validation is needed.

The last aspect of LiPON's stability against LNMO correlates with its morphological characteristics. LiPON is known for its amorphous structure and dense, pinhole-free feature. Such merits of LiPON give rise to its even coverage over LNMO after deposition and facilitates a uniform overlithiation on an LNMO surface, namely a conformal interface. In the case of a liquid electrolyte system, the CEI tends to be non-uniform in both composition and thickness, due to the hardly controllable mass transfer in a liquid environment, which diminishes the passivation effect of such interfaces. Compared to studies with other solid electrolytes (i.e., $Li_7La_3Zr_2O_{12}$ and $Li_6PS_5Cl$, etc.), roughness at the pelletized cathode/electrolyte interface can incur inhomogeneity in terms of contact and pressure, resulting in interface non-uniformity[72,73]. Given the compatibility of LiPON against varieties of cathodes, an artificial LiPON layer deposited on a cathode surface could serve as a simple protection strategy to alleviate the interfacial reaction between a less electrochemically stable solid electrolyte and highly oxidative cathode material. Alternatively, employing methodologies for producing glassy material (e.g., via fast quenching) to amorphize conventionally crystalline solid electrolytes through the creation of defects could well steer the morphological features and enhance interfacial stability.

It should be noted that the absence of a carbon conductive agent also helps to achieve the stable cycling performance. Although carbon additives improve electronic conductivity of a cathode or an anode composite electrode, past researches has indicated that carbon additives can accelerate the decomposition of an electrolyte and form CEI on a cathode's surface[74,75] and SEI on an anode's surface[76], causing capacity degradation of cells. The high electronic conductivity of the carbon also creates an electronic pathway for electrolyte redox reactions to occur. Therefore, it is critical to avoid carbon or minimize its surface area to suppress the kinetics of the electrolyte decomposition.

To better illustrate the interface configuration between LNMO and LiPON, a schematic is shown in **Figure 5**. Examinations of an LNMO/LiPON interface using NDP, DFT and cryo-EM yields intriguing results that are closely related to the stability of the materials' interface. An ideal cathode electrolyte interface requires the electrolyte to remain either chemically or electrochemically stable against the cathode and mechanically robust. An as-formed CEI ought to be uniform and conformal, consisting of species that are electronical insulating and ionically conductive to prevent further decomposition of electrolyte. Finally, the cathode should be able to retain the crystal structure after extended cycling.

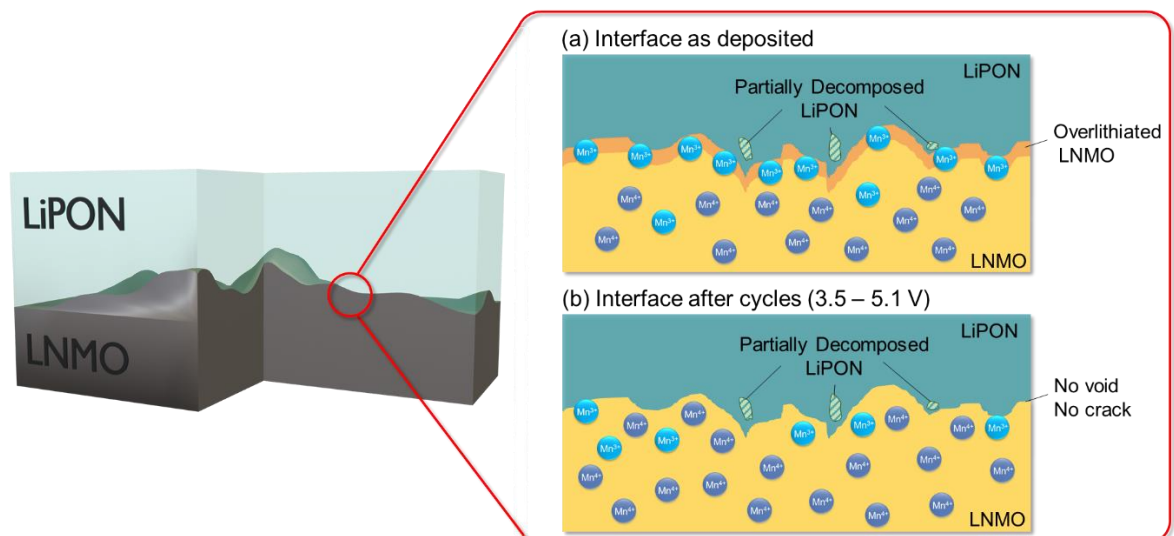

Figure 5. Schematic illustration of microscopic LNMO/LiPON interface At the LNMO/LiPON interface (left), microscopically LNMO/LiPON contact without void and cracks both (a) as deposited and (b) after cycling between 3.5 V - 5.1 V. As deposited interface in (a), overlithiated regions is likely to exist in LNMO, where Mn oxidation state is reduced to 3+. After cycling (b), such area disappears and $Mn^{4+}$ is dominant in overall LNMO. In both interface, partial decomposition of LiPON is observed. Not to scale.

**Conclusions**

New insights into the interface between LNMO cathode and LiPON solid electrolyte have been gained by combining targeted spectroscopic, microscopic, and computational tools. An overlithiation layer of LNMO at the LNMO/LiPON interface due to LiPON deposition is revealed by NDP, which accounts for the excess capacity during the $1^{st}$ charge of LNMO. Overlithiated LNMO exhibits a Mn oxidation state change from 4+ to 3+ as confirmed by both DFT prediction and cryo-STEM/EELS examination. The cathode structure is intact under the overlithiation and is not significantly impacted after electrochemical activation. Cryo-TEM further shows a LNMO/LiPON interface with intimate contact that is free of voids and cracks over the course of 500 cycles. STEM/EELS demonstrates that major elements across the interface keep their original chemical bonding environment. The observed conformal and stable interface contributes to the electrochemical stability of LiPON against LNMO. The knowledge gained from additive-free electrodes and its close contact to mechanically and electrochemically compatible SSE demonstrate that the long-term cycling stability of 5 V-class cathode materials is achievable with dedicated interface engineering.

**Experimental Procedures**
*Thin film sample preparation*
LNMO thin-films were deposited on platinum-coated alumina substrate (Valley Design) by pulsed laser deposition (PLD) system (Excel Instruments PLD STD-12 chamber and 248 nm KrF Lambda Physik-Pro 210 excimer laser) with a laser energy fluence of ≈ 2.0 J cm$^{-2}$ and 24000 pulses at a frequency of 10 Hz. The substrate temperature was heated to 600 °C, and partial pressure of $O_2$ was controlled at 200 mTorr (1 mTorr = 133.322 Pa, SI unit) during the deposition. The LNMO target used for PLD was prepared using LNMO powder (NEI corporation, USA). 12 g of LNMO powder and 0.47 g of LiOH (Sigma-Aldrich) were used to achieve a 30% Li excess LNMO target. The materials were ball milled and pelletized onto a 28.6 mm dye press at a pressure of 10 Mg for 10 min. Following this step, the pellet was sintered at 900 °C for 2 h with a heating ramp rate of 3 °C min$^{-1}$. The resulting target surface was polished by sandpaper (Grit 320 and 600, Aluminum oxide, Norton abrasives) before every deposition.

LiPON thin films were deposited by radio-frequency (RF) sputtering. A $Li_3PO_4$ target that was ≈ 50 mm in diameter (Plasmaterials, Inc.) was used as a sputtering target. The sputtering power was set at 50 kg·m$^2$·s$^{-3}$ (W). Nitrogen gas (Matheson, ultra-high purity grade) and the partial pressure at 15 mTorr for deposition.

LNMO was deposited to a thickness of ≈ 650 nm and an active area of 4.9 mm$^2$, corresponding to an active material loading of 0.03 mAh cm$^{-2}$ for the full cell fabrication. LiPON was then sputtered on LNMO to a thickness of 1 μm. A Li metal anode and Cu current collector were then deposited on to the LiPON by thermal evaporation (LC Technology Solutions Inc.) under a base pressure below 3×10$^{-11}$ mTorr. The average deposition rate of Cu and Li were controlled at 0.1 nm s$^{-1}$ and 0.15 nm s$^{-1}$, respectively. Deposition was monitored by a quartz crystal microbalance. The thickness of Li metal anode was 570 nm and corresponds to a 203% excess capacity compared to that of the cathode. The thickness of the LNMO and LiPON were controlled at 2.6 μm and 2.2 μm thick, respectively, for the NDP measured samples.

*Grazing incidence angle X-ray diffraction (GIXRD)*
XRD pattern of the fabricated thin film was taken by Rigaku Smartlab X-ray diffractometer with Cu Kα source ($\lambda$ = 1.5406 Å; 1 Å = 0.1 nm, SI Units) with a working voltage and current of 40 kV and 44 mA, respectively, and a scan step size of 0.04°. The scan speed was 0.12° min$^{-1}$, and the scan range was from 15° to 80°.

*Liquid cell fabrication*
The LNMO thin films were also tested with liquid electrolyte in a coin cell. An LNMO thin film deposited on a Pt-coated alumina substrate (1 cm$^2$ surface area, cathode) was coupled with ≈ 100 μL of 1M $LiPF_6$ in EC:EMC (3:7 wt%) electrolyte and a Li metal chip as an anode. The coin cell consisted of a CR2032 type casing and one piece of a Celard 2325 separator. The cell was cycled between 3.5 V and 4.85 V with a 10 μA constant current.

*Electrochemical measurement*
Two LNMO/LiPON/Li full cells were cycled with a Biologic SP-200 potentiostat. For the first cell, the voltage range of the cycling was set at 3.5 V - 5.1 V and the applied current was 150 nA and 3.0 μA. This is equivalent to a ≈ C/10 and ≈ 5C charging rate, respectively. The cell was cycled at C/10 during the first 2 cycles and the last 3 cycles (1st, 2nd, and 533rd to 535th cycle) and cycled at 5C for the remaining cycles. The electrochemical cycling data of this cell has already been published in our past study[49] and its characterization by (S)TEM were conducted, shown in this paper. The other full cell was cycled for 600 cycles in the same voltage range with the current at 100 nA for 1st, 2nd, and 600th cycles and 1.7 µA for the remaining cycles, which are almost equivalent to ≈ C/7 and ≈ 4C charging rate, respectively. The second cell was fabricated for the purpose of confirming reproducibility of the cell performance, demonstrated in this paper. The areal capacity of the cells was calculated based on the assumption that only part of LNMO above Pt current collector (≈ 3 mm in diameter) was involved in the electrochemical reaction.

*Neutron depth profiling and fitting*
Neutron depth profiling (NDP) data was collected at the National Institute of Standards and Technology (NIST) Center for Neutron Research (NCNR) at the end position of the cold Neutron Guide 5. The $^6$Li atoms in the sample were of interest for these experiments and measured through detection of the $^3$H$^+$ (triton) charged particle products from the $n,^6$Li reaction:

$$^6Li + n_{cold} \rightarrow {}^4He^{2+}(\alpha, 2055.55\ keV) + {}^3H^+({}^3H, 2727.92\ keV) \qquad \text{Equation 1}$$

Following reaction $^4$He$^{2+}$ and $^3$H$^+$ particles are promptly produced and immediately begin to lose energy due their interactions with the nuclear and electronic properties of the sample material. However, the $^4$He particles were blocked from the detector in this experiment by the polyimide cover that was added to the surface of the sample to protect the battery material from the ambient atmosphere. Only the $^3$H$^+$ particles were analyzed for all the samples measured.

The sample was mounted behind a 0.5 mm thick Teflon sheet with a ≈ 3.0 mm circular aperture. This aperture was fixed to an Al support frame and placed facing the primary transmission-type silicon surface barrier detector (Ametek) inside the NDP chamber. The energy spectra of the detected particles were collected and transmitted to a Lynx$^{TM}$ Digital Signal Analyzer (Canberra) with a setting of 4092 channels. Data were acquired for ≈ 5 h per sample. Each sample area was irradiated at a neutron fluence rate of ≈ 1.2×10$^9$ neutrons cm$^{-1}$ sec$^{-1}$. A high-vacuum chamber was used for the measurements of the sample, background profiles (Teflon, Si wafer), and a $^{10}$B concentration reference material (in house). Li atom concentrations were calculated using the natural abundance of $^6$Li for sample and $^{10}$B as a reference. Li concentration was calculated by the following equation:

$$D_{Li} = \frac{D_{10B} \times \sigma_{10B}[barns]}{n_{6Li} \times C_{10B}[counts\ sec^{-1}] \times \sigma_{6Li}[barns]} \times C_{6Li}[counts\ sec^{-1}] \qquad \text{Equation 2}$$

$D_i$ is the areal concentration of $^{10}$B, $^6$Li, or Li (*i*) in *i* atoms cm$^{-2}$, $\sigma_i$ is the thermal neutron cross-section for the isotope *i* (1 barn = 1x10$^{-28}$ m$^2$), $n_{6Li}$ is the natural abundance of $^6$Li, and $C_i$ is the normalized particle counts detected in the measurement. The calculated data was binned according to energy resolution of the NIST NDP system (≈ 22 keV for a $^3$H at 2727 keV). More details of the data processing can be found in the SI.

The uncertainties of both LNMO/LiPON and LNMO are <6% and reported to 1 sigma, which are estimated from the propagation of the experimental counting statistics from the sample, reference, and background materials.

*First-principles calculations*
LNMO at various charged/discharged states were studied by density functional theory (DFT) with the Vienna *Ab initio* Simulation Package (VASP). Periodic plane-wave DFT+U static calculations were performed for the LNMO bulk structure. Supercell models, $Li_8Ni_4Mn_{24}O_{32}$, were used as $Li_xNi_{0.5}Mn_{1.5}O_4$ at $x = 1$. To simulate Li removal from $x = 1$ to 0 a corresponding number of Li atoms were removed at each state. Li atoms were inserted into tetrahedral sites between octahedral sites containing Ni or Mn atoms to simulate the overlithiation of the cathode material. A 3×3×3 k-point mesh and an energy cutoff of 520 eV were employed through the calculation. $U_{eff}$ values were chosen as 5.96 eV and 4.5 eV for the +U augmented treatment of Mn and Ni 3d orbitals, respectively. The initial MAGMOM parameters were set as follows: Li*(0), Ni*(-2), Mn*(+4), O*(0). Partial density of states (PDOS) at pristine state ($x =1$) and Overlithiated state ($x = 2$) were extracted from DOSCAR and analyzed by wxDragon software. PDOS plots were smoothed by a 10-point adjacent average function.

*Cryogenic focused ion beam/scanning electron microscopy (cryo-FIB/SEM)*
A FEI Scios DualBeam FIB/SEM with a cooling stage was used to prepare the TEM samples of pristine LNMO/LiPON and cycled LNMO/LiPON/Li samples. The operating voltage of the electron beam was 5 kV and emission current of the beam was 50 pA. These setting were used to mitigate potential beam damage on Li and LiPON. A Ga ion beam source was used to mill and thin the sample with an operating ion beam voltage of 30 kV. Emission currents of the ion beam were selected depending on purposes: 10 pA for ion beam imaging, 0.1 nA for cross-section surface cleaning and lamella thinning, and 3 nA for pattern milling. The sample stage temperature was maintained at -185 °C during pattern milling, cross-section cleaning, and lamella thinning processes to preserve the Li metal and LiPON components. A cryo-liftout methodology was applied for the TEM sample lift-out process where sample materials were redeposited between the lamella and the probe for connection. Detailed procedures with illustration were described in our past study[49].

*Cryogenic transmission electron microscopy (cryo-TEM)*
The prepared lamellas were transferred from the FIB chamber though an air-free quick loader and stored in an Ar-filled glovebox. Cryo-HRTEM images were collected on a JEOL JEM-2100F TEM at 200 kV with a Gatan Oneview camera. Cryo-STEM/ EELS results were collected on a JEOL JEM-ARM300CF TEM at 300 kV. A TEM cryo-holder (Gatan 626 cryo-transfer holder) was used to load the samples where TEM grids were immersed in liquid nitrogen and then mounted onto the holder via a cryo-transfer workstation[49]. Uncertainty of the energy loss in EELS spectra are 1 eV, which is attributed to the energy resolution of the detector.

**Acknowledgements**
The authors gratefully acknowledge funding support from the US Department of Energy, Office of Basic Energy Sciences, under award number DE-SC0002357. FIB/SEM was performed at the San Diego Nanotechnology Infrastructure (SDNI), a member of the National Nanotechnology

Coordinated Infrastructure, which is supported by the National Science Foundation (Grant ECCS-2025752). TEM was performed at the UC Irvine Materials Research Institute (IMRI). This work also used the Extreme Science and Engineering Discovery Environment (XSEDE), which is supported by National Science Foundation grant number ACI-1548562. The authors would also like to thank the National Institute of Standards and Technology (NIST) for the use of neutron depth profiling facility at the NIST Center for Neutron Research (NCNR). Certain commercial equipment, instruments, or materials (or suppliers, or software) are identified in this paper to foster understanding. Such identification does not imply recommendation or endorsement by the National Institute of Standards and Technology, nor does it imply that the materials or equipment identified are necessarily the best available for the purpose.